\documentclass[prb,twocolumn,superscriptaddress,showpacs]{revtex4}
\usepackage[dvips]{graphicx}
\usepackage{latexsym}
\usepackage{bm}
\usepackage{psfrag}
\usepackage{wasysym}
\usepackage{color}

\newcommand{\fig}[2]{\includegraphics[width=#1]{#2}}
\definecolor{Blue}{rgb}{0.00, 0.00, 1.00}
\definecolor{Red}{rgb}{1.00, 0.00, 0.00}

\begin{document}

\title{ Quantum Spin Hall Effect and Topologically Invariant Chern
Numbers}
\author{D. N. Sheng}
\affiliation{Department of Physics and Astronomy, California State
University, Northridge, CA 91330}
\author{Z. Y. Weng}
\affiliation{Center for Advanced Study, Tsinghua University,
Beijing 100084, China}
\author{L. Sheng}
\affiliation{Department of Physics and Texas Center for
Superconductivity, University of Houston, Houston, TX 77204}
\author{F. D. M. Haldane}
\affiliation{Department of Physics, Princeton University,
Princeton, NJ 08544}

\begin{abstract}
We present a topological description of quantum spin Hall effect
(QSHE) in a two-dimensional electron system on honeycomb lattice
with both intrinsic and Rashba spin-orbit couplings. We show that
the topology of the band insulator can be characterized by a
$2\times 2$ traceless matrix of first Chern integers. The
nontrivial QSHE phase is identified by the nonzero diagonal matrix
elements of the Chern number matrix (CNM). A spin Chern number is
derived from the CNM, which is conserved in the presence of finite
disorder scattering and spin nonconserving Rashba coupling. By
using the Laughlin's gedanken experiment, we numerically calculate
the spin polarization and spin transfer rate of the conducting
edge states, and determine a phase diagram for the QSHE.
\end{abstract}

\mbox{}\\

\maketitle

Topological quantities are fundamentally important in
characterizing a new kind of order---topological order~\cite{wen},
which cannot be described in the paradigm of a Landau symmetry
breaking theory. For two-dimensional (2D) electron systems in
perpendicular magnetic fields, they are manifested in the
transverse electrical transport  in integer and fractional quantum
Hall effect (IQHE/FQHE) states~\cite{qheb}. It was first revealed
by Thouless {\emph et. al.}~\cite{thouless} that each IQHE state
is associated with a topologically invariant integer known as
first Chern number, which precisely equals the Hall conductance in
units of $e^{2}/h$. The exact quantization of the Hall conductance
can also be formulated in terms of a 2D band-structure Berry
phase~\cite{thouless,berry,niu,haldane}, which remains an integral
invariant till the band energy gap (or more precisely the mobility
gap~\cite{donna1} in the presence of disorder) collapses.

While the conventional IQHE is usually associated with strong
magnetic fields, Haldane~\cite{haldane88} has explicitly shown
that it can actually occur in the absence of magnetic field in
band insulators with graphene-like band structure. The
one-component Haldane's model explicitly breaks time-reversal
symmetry, resulting in a condensed-matter realization of a parity
symmetry anomaly with chiral edge states at the boundary of the
sample. In realistic electron systems, however, the coupled spin
degrees of freedom can recover the time reversal symmetry by
forming Kramers degenerate states, which belong to the
universality class of zero charge Chern number as the total Berry
curvature of the occupied energy band of both spins sums to zero.

This class of insulator has been recently
found~\cite{kane,lisheng} to possess a dissipationless quantum
spin Hall effect (QSHE)~\cite{zhang}, which is distinct  from the
intrinsic spin Hall effect (SHE) in the metallic
systems~\cite{intrinsic}. The QSHE has been shown to be robust
against disorder scattering and other perturbation
effects~\cite{kane,lisheng}. Whether there exists an underlying
topological invariant \textquotedblleft
protecting\textquotedblright\ the QSHE is a very important issue
for both fundamental understanding and potential applications of
the QSHE. While the previously proposed~\cite{kane,moore}
$Z_{2}$~classification of the QSHE suggests that the  conducting
edge states are protected by time-reversal symmetry, it does not
distinguish between two QSHE states with spin Hall conductance
(SHC) of opposite signs. Thus it remains an open issue if the QSHE
states can be classified by a more definite topological quantum
number similar to the Chern number for the conventional IQHE.

In this Letter, we clarify the topological nature of the QSHE in
graphene model with spin-orbit coupling. We present numerical
evidence that a pair of nonchiral conducting edge states are
responsible for the dissipationless spin current in the open
system with disorder scattering. The spin transfer and spin
accumulation are calculated, which increase linearly with the flux
insertion in the Laughlin's gedanken experiment\cite{laugh}. We
further establish the relation between the conducting edge states,
the topological invariant Chern number matrix (CNM) and a new
conserved spin Chern number for the bulk system with electron
Fermi energy lying inside the band gap. The nontrivial spin Chern
number distinguishes a QSHE state from an ordinary band insulating
state, and is responsible for the robust QSHC.

We begin with the 2D honeycomb lattice
model~\cite{haldane88,kane,lisheng}, which is relevant to the 2D
electrons in a single-atomic layer graphene system\cite{graph}:
\begin{eqnarray}
H_{0} &=&-t\sum\limits_{\langle ij\rangle
}c_{i}^{\dagger}c_{j}+\frac{2i}{ \sqrt{3}}V_{{\tiny
SO}}\sum\limits_{\langle\langle ij\rangle \rangle }c_{i}^{\dagger
}\mathbf{\sigma }\cdot\left( \mathbf{d}_{kj}\times \mathbf{d}
_{ik}\right) c_{j}  \nonumber \\
&+&iV_{R}\sum\limits_{\langle ij\rangle}c_{i}^{\dagger }\hat{
\mathbf{z}}\cdot \left( \mathbf{\sigma}\times
\mathbf{d}_{ij}\right) c_{j}+\sum_{i}w_{i}c_{i}^{\dagger}c_{i}\ ,
\label{HAMIL}
\end{eqnarray}
where $c_{i}^{\dagger }=(c_{i\uparrow }^{\dagger },c_{i\downarrow
}^{\dagger })$ are electron creation operators, and
${\mbox{\boldmath$\sigma$}}$ is the Pauli matrix. The first term
is the usual nearest neighbor hopping term, and the second term is
an intrinsic SO coupling preserving the lattice symmetries with
$i$ and $j$ as two next nearest neighbor sites and $k$ their
unique common nearest neighbor. Here, the vector $\mathbf{d}_{ik}$
points from $k$ to $i$, with the distance between two nearest
neighbor sites taken to be unity. The third term stands for the
Rashba SO coupling with strength $V_{R}$, and the last term
represents an on-site random disorder with strength $ \left\vert
w_{i}\right\vert \leq W/2$.

In the absence of the Rashba SO coupling $V_{\tiny R}=0$, the
model Eq.\ (\ref{HAMIL}) reduces to a two-component Haldane's
model~\cite{haldane88}, which exhibits a quantized SHC $\sigma
_{sH}=\pm 2 \frac {e}{4 \pi}$ with sign depending on the sign of
$V_{SO}$. Such a QSHE has been shown to persist into a finite
range of $V_{ R}$~\cite{kane} and disorder strength
$W$~\cite{lisheng}, so long as the bulk band gap remains finite.
In the following, we will perform systematic numerical simulations
designed to reveal the topological characteristics of such a
nontrivial band insulator.

\begin{figure}[tbp]
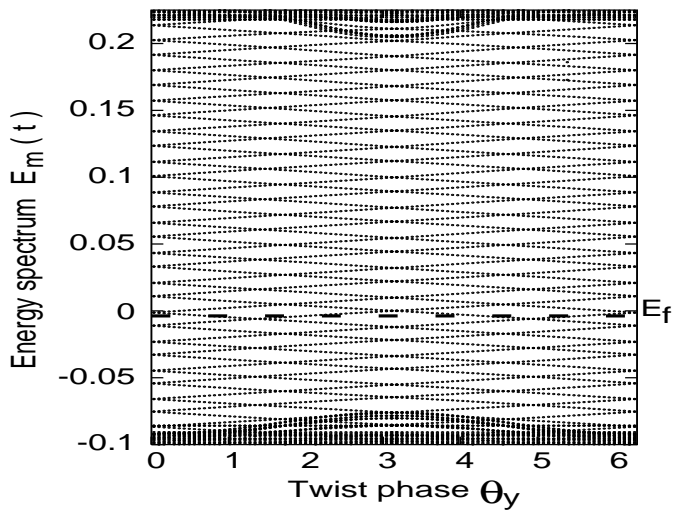

\begin{center}
\vskip-0.7cm \hspace*{-1.3cm} \fig{3.8in}{sfig1.ps} \vskip-2mm
\vskip-8mm
\end{center}
\vskip -0.4cm \caption{The evolution of edge levels inside the
bulk gap with twist
boundary phase $\protect\theta_y$ in a cylindric 2D sample of $%
N=480\times 240$, $V_{SO}=0.05t$, $V_{R}=0.05t$ and $W=1.0t$.}
\label{fig:fig1} \vskip -0.45cm
\end{figure}

\textit{Gauge experiment and edge states free from backward
scattering---}Let us first study a rectangular geometry of
honeycomb lattice consisting of $N_{y}$ zigzag chains with $N_{x}$
atom sites on each chain. A twisted boundary condition is
imposed~\cite{donnarashba} in the $\hat{y}$-direction, and an open
boundary condition is used in the $\hat{x}$-direction. The system
is also equivalent to a cylindric geometry threaded by a flux
$\Phi =\frac{\theta_{y}}{2\pi}$ in the axial direction
($\hat{x}$-direction), similar to that used in the Laughlin's
gedanken experiment~\cite{laugh} for IQHE. The calculated energy
levels for $V_{SO}=0.05t$, $V_R=0.05t$, $W=1.0t$ and system size
$N\equiv N_{x}\times N_{y}=480\times 240$ are plotted in Fig.\
\ref{fig:fig1}. We see that some new states emerge within the bulk
energy gap $-0.1t<E_{m}<0.2t$ with the density of states much
smaller than that of the bulk states outside the energy gap. These
mid-gap states can be identified as edge states with their
wavefunctions localized within a few lattice constants at the two
open boundaries in the $\hat{x}$-direction and extended along the
$\hat{y}$-direction.

 The edge states in Fig.\ \ref{fig:fig1}
 continually evolve and cross each other with
increasing $\theta _{y}$. Similar to IQHE, the level crossing of
the edge states in Fig.\ \ref{fig:fig1} clearly demonstrates the
absence of backward scattering, and thus each state will not
simply evolve back to its original state after the insertion of
one flux quantum $\Phi =1$ (or $\theta _{y}=2\pi$), which leads to
an adiabatic transverse transport of electrons from one edge to
the other\cite{donnarashba}. For a Fermi energy $E_f$ inside the band gap, as
indicated by the dashed line in Fig.\ \ref{fig:fig1}, there are
always a pair of edge states crossing the $E_f$ line and evolving
out of the Fermi sea. They can be identified in the real space as
moving oppositely, either from the left edge to the right one or
vice versa. Thus these ideal conducting channels are nonchiral,
unlike the IQHE, and do not contribute to a net transverse
electric charge transport.

\begin{figure}[tbp]
\begin{center}
\vskip-0.8cm \hspace*{-0.50cm} \fig{4.1in}{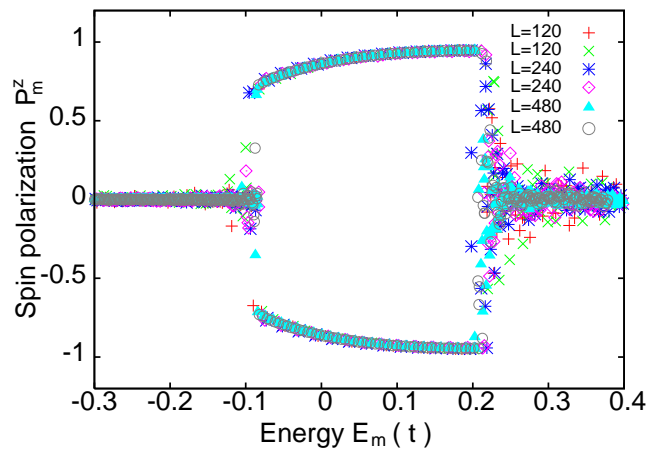} \vskip-2mm
\vskip-8mm
\end{center}
\vskip-0.4cm \caption{Spin polarization $P^{z}_m$ defined in the
text (in units of $\hbar/2$) as a function of eigen energy $E_{m}$
for two random disorder configurations (we have checked over 100  
configurations) of strength $W=1.0t$ at different system sizes 
$N=120\times 120-480\times 480$ (with $N_{x}=N_{y}=L$). 
$V_{SO}$ and $V_{R}$ are the same as in Fig.\ 1.
} \label{fig:fig2} \vskip-0.35cm \vskip-0.1cm
\end{figure}

\textit{Robust spin polarization carried by the edge states}---We
then explicitly compute the spin polarization
$P^{z}_m=\frac{\hbar}{2}\langle m|\sum_{i}c_{i}^{+}\sigma
^{z}c_{i}|m\rangle$ along the $\hat{z}$-direction with $|m\rangle
$ as the $m$-th single-particle eigenstate. Spin polarizations
along the $\hat{x}$- and $\hat{y}$-directions are found much
smaller. At $\theta _{y}=0$, all the eigenstates form Kramers
degenerate pairs, and each pair carry totally zero spin due to
time-reversal symmetry. We find that upon insertion of the flux
($\theta_{y}\neq 0$), a nonzero spin polarization is quickly
developed for each edge state. Fig.\ \ref{fig:fig2} shows the spin
polarization of each eigenstate as a function of eigenenergy
$E_{m}$  at  $\theta_y=2\pi/48$.  The spin polarization
is finite only for the edge states within the energy gap and
insensitive to disorder configurations and sample sizes, as all
the data points nicely collapse together. We find that $P^{z}_m$
is mainly affected by the Rashba coupling $V_{R}$ and decreases
monotonically with increasing $V_{R}$.

\begin{figure}[tbp]
\begin{center}
\vskip-0.7cm \hspace*{-0.50cm} \fig{3.6in}{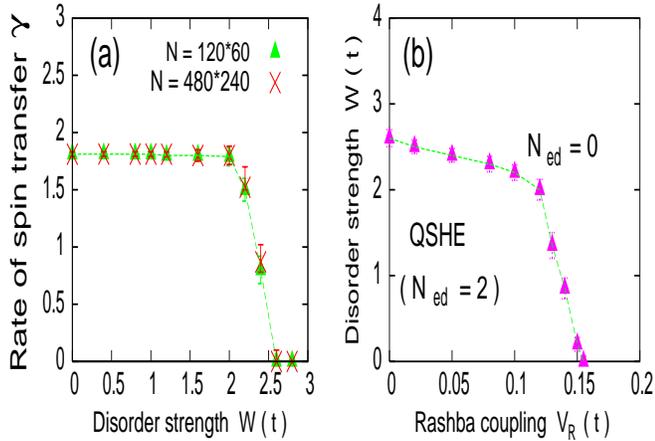} \vskip-6mm
\vskip-8mm
\end{center}
\vskip-0.2cm \caption{(a) Spin transfer rate $\gamma=\frac
{\Delta\langle S^z\rangle_{\mbox{\tiny edge}}}{\Delta\Phi}$ (in
units of $e/4\pi$)
 vs. disorder strength $W$
at $V_{SO}=0.05t$ and $V_{R}=0.05t$ averaged over $500$ disorder
configurations; (b) The phase boundary for
the QSHE determined from the spin transfer, where the transition
happens with a change of the numbers of edge channels from
$N_{ed}=2$ to
$N_{ed}=0$%
.} \label{fig:fig3} \vskip-0.3cm
\end{figure}

\textit{Spin transfer rate and phase diagram}---Now we can
determine the spin transfer by the edge states with the adiabatic
insertion of the magnetic flux. There are always a pair of states
crossing the $E_{f}=0$ line from below, as shown in Fig.\
\ref{fig:fig1}. One state carries a positive $P_m^{z}$, which
moves to the right edge of the sample, and the other with negative
$P_m^{z}$ moves to the left edge, as determined from the
wavefunctions. Thus after the insertion of one flux quantum
($\theta _{y}=0\rightarrow 2\pi$), a net spin transfer occurs,
which results in spin accumulations at the two open edges of the
sample. We can determine the average spin transfer rate $\gamma$
by using the relation $\gamma=\frac{\Delta \left\langle
S^{z}\right\rangle_{\mathrm{edge}}}{\Delta \Phi}
=\int_{0}^{2\pi}d\theta_{y}\frac{d\left\langle S^{z}\right\rangle
_{\mathrm{edge}}}{d\theta_{y}}$ with $\langle
S^{z}\rangle_{\mathrm{edge}%
}=\sum_{m}^{\prime }P_m^{z}$, where the summation runs over only
the edge states at the right edge. As shown in Fig.\
\ref{fig:fig3}a, the spin transfer rate remains about $1.8$ (in
units of $\frac{e}{4\pi}$) till a critical disorder strength
$W_{c}\sim 2.4t$ is reached, beyond which $\gamma$ vanishes
quickly. By analogy with IQHE~\cite{laugh}, the spin transfer rate
corresponds to a finite SHC of value around $1.8\frac{e}{4\pi}$,
which is in good agreement with the value obtained by using the
Landauer-B\"{u}ttiker formula~\cite{lisheng}.

The robust and dissipationless edge states thus clearly
distinguishes the QSHE in the present system from the intrinsic
SHE in other metallic systems. In Fig.\ \ref{fig:fig3}b, a phase
diagram in the parameter plane of the Rashba coupling $V_{\tiny
R}$ and disorder strength $W$ is obtained, where the phase
boundary of the QSHE regime is determined by numerical calculation
of the above spin transfer rate. We find that such a phase
boundary is well correlated with the collapse of the bulk energy
gap, although the precise position of the latter is harder to
pinpoint at finite $W$.

\textit{Topologically invariant Chern number matrix---}At this
stage, we have firmly established the connection between the QSHE
in the model Eq.\ (\ref{HAMIL}) and the presence of edge states
within the bulk band gap. In IQHE, it is well known that edge
states are closely related to the bulk topological quantity, i.e.,
the first Chern number. In the following, we examine whether there
also exists a topological characterization of the bulk states,
which is responsible for the existence of the edge states.

Generally, in diagonalizing the Hamiltonian Eq.\ (\ref{HAMIL}) one
may introduce a generalized boundary condition~\cite{niu,donna2}
for a 2D
many-body wavefunction: $\Phi (...,\mathbf{r}_{i_\alpha }+\mathbf{L}%
_{j},...)=e^{i\theta _{j}^{\alpha }}\Phi
(...,\mathbf{r}_{i_{\alpha }},...)$ , where $j=x,y$, and the
system length vector $\mathbf{L}_{x}= \frac
{N_{x}}2\mathbf{a}_{1}$ and $\mathbf{L}_{y}=N_{y}\mathbf{a}_{2}$,
with $\mathbf{a}_{1}$ and $ \mathbf{a}_{2}$ as two primitive
vectors of the Bravais lattice. The twisted boundary condition is
represented by $0\leq \theta _{j}^{\alpha }<2\pi$, where
$\alpha=\uparrow$ and $\downarrow$ denotes the spin index. Through
a unitary transformation $\Psi =\exp [-i\sum_{\alpha
}\sum_{i_{\alpha }}(\frac{\theta
_{x}^{\alpha }}{L_{x}}x_{i_\alpha }+\frac{\theta_y^{\alpha }}{L_{y}}%
y_{i_\alpha })]\Phi $, where the summation runs over all electrons
of both spins, $\Psi$ becomes periodic on a torus. One can then
define the topological Chern numbers as~\cite{thouless,donna2}:
\begin{equation}
C^{\alpha ,\beta }={\frac{i}{4\pi }}\int \int d\theta _{x}^{\alpha
}d\theta _{y}^{\beta }\left[ \Bigl\langle {\ \frac{\partial \Psi
}{\partial \theta _{x}^{\alpha }}\Bigl|{\frac{\partial \Psi
}{\partial \theta _{y}^{\beta
}}}%
\Bigr\rangle -\Bigl\langle \frac{\partial \Psi }{\partial \theta
_{y}^{\beta }}\Bigr|}\frac{\partial \Psi }{\partial \theta
_{x}^{\alpha
}}\Bigr\rangle %
\right]   \label{cnm}
\end{equation}
where the area integration is over a unit cell $0\leq \theta
_{x}^{\alpha }$ , $\theta _{y}^{\beta }\leq 2\pi $. With
$\alpha,\beta=\uparrow,\downarrow$, $C^{\alpha ,\beta }$ form a
$2\times 2$ CNM~\cite{donna2}.

The many-body wavefunction $\Psi(\theta )$ is necessarily a smooth
function of the boundary phase $\theta =\{\theta _{x}^{\alpha
},\theta _{y}^{\beta }\}$ when the energy gap remains. One can
then prove the exact quantization of $C^{\alpha,\beta }$ by
strictly following the argument of Thouless $et$
$al.$~\cite{thouless}. Each topologically invariant matrix element
$C^{\alpha,\beta }$ should remain unchanged until the energy gap
collapses. Without breaking the time reversal symmetry, the trace
of the CNM should always vanish. Consider the simple case without
Rashba coupling term ($V_{R}=0$), the only nonzero matrix elements
are the diagonal ones: $C^{\alpha,\alpha}=\eta_{\alpha}$ with
$\eta_{\uparrow}=-\eta_{\downarrow}=1$ for $V_{SO}>0$, which
change sign with $V_{SO}$. They describe the IQHE in decoupled
subsystems of $\alpha=\uparrow $ and $\downarrow$. While the total
charge Chern number $C_{c}\equiv \sum_{\alpha ,\beta }C^{\alpha
,\beta }$ (corresponding to the total charge Hall conductance of
the system)~\cite{donna2} cancels out, the total spin-related
Chern number is quantized to $C_{sc}\equiv \sum_{\alpha ,\beta
}\alpha C^{\alpha ,\beta }=2$, which represents the spin
Hall response when a common electric field for both spin
components is imposed along the $\hat{y}$-direction. In this spin
decoupled limit, the quantized spin Chern number $C_{sc}$ is
associated with the SHC by $\sigma _{sH}=C_{sc}=2$ in units of
$\frac{e}{4\pi}$.

Now we turn on the Rashba coupling and numerically determine
$C_{sc}$. This can be done either by calculating each $C^{\alpha
,\beta }$ first or carrying out the integration in Eq.\
(\ref{cnm}) for opposite boundary condition along the
$\hat{x}$-direction (spin twist $\theta _{x}^{\uparrow}=-\theta
_{x}^{\downarrow}=\theta _{x}$) and common one along the
$\hat{y}$-direction ($%
\theta _{y}^{\uparrow}=\theta _{y}^{\downarrow}=\theta _{y}$) to
directly obtain $C_{sc}$. We choose the latter method in the
following. The unit cell of the boundary phases is divided into
$N_{\mathrm{mesh}}=400$ to $14400$ mesh points such that the
integration in Eq. (\ref{cnm}) is replaced by the sum of the solid
angle $\Omega _{j}$: $C_{sc}=\sum_{j}\frac{\Omega _{j}}{2\pi} $.
Here, $\Omega _{j}=\arg \prod_{i}\langle \Psi _{j_{i}}|\Psi
_{j_{i+1}}\rangle $ where $i=1-4\,$ (with $j_{5}\equiv j_{1})$
denote four mesh points at the $j$-th square of mesh patches in
the $\theta $-space. Numerically it is verified that the computed
$C_{sc}$ is well converged and insensitive to mesh sizes. We find
that $C_{sc}$ always remains quantized at $C_{sc}=2$ before the
band gap collapses at a critical Rashba coupling, say,
$V_{R}^{c}=0.33t$ if $V_{SO}$ is fixed at $0.1t$. Such a
topologically nontrivial phase also survives for very small
$V_{SO}$, so long as $V_{R}$ is not much larger than $V_{SO}$.

\begin{figure}[tbp]
\begin{center}
\vskip-1.9cm \hspace*{-1.1cm} \fig{3.8in}{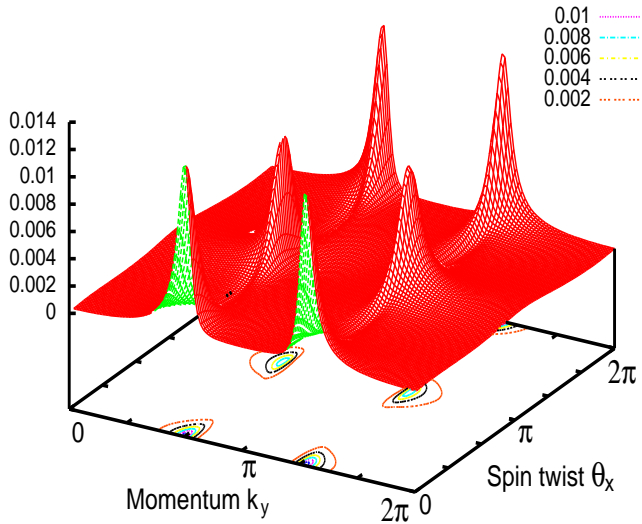} \vskip-2mm
\vskip-8mm
\end{center}
\vskip-0.5cm \caption{Solid angle $\Omega_{j}$ vs.
spin twist $\protect\theta _x$ and momentum $k_y$, which is
meshed into $N_{mesh}=120 \times 120$ points for a pure system with $%
N=240\times 240$ at $V_{SO}=0.1t$ and $V_R=0.3t$. The total Berry
phase $ \sum_{j=1}^{N_{mesh}} \Omega_j=4\protect\pi$ and thus
$C_{sc}=2$.} \label{fig:fig4} \vskip-0.5cm
\end{figure}

The total spin Chern number $C_{sc}$ can be also expressed as the
sum of the Berry phase of each eigenstate below the Fermi energy.
Since along the $\hat{y}$-direction, a common boundary condition
is imposed, which does not break the translational invariance of
the system along this direction, one can calculate the
contribution to $C_{sc}$ of the eigenstates for a given momentum
$k_{y}$ and formulate $C_{sc}$ as an area integral over $k_{y}$
and $\theta_{x}$. Interestingly, it is found that all eigenstates
contribute more or less uniformly to $C_{sc}$ at small $V_{R}$,
which is responsible for the robustness of the topological
invariant $C_{sc}$. On the contrary, at a larger $V_{R}=0.30t$,
the energy gap is about to collapse, and correspondingly the
nonzero $\Omega _{j}$ becomes quite localized and concentrated at
four points in the phase space $(0,\pm k_{y}^{\ast })$
and $%
(\pi ,\pm k_{y}^{\ast })$, as shown in Fig.\ \ref{fig:fig4}, where
$k_{y}^{\ast }=\frac{2\pi}{3}$ (it appears that there are six
peaks simply because of the periodicity in $\theta _{x}$). As a
matter of fact, we find that each peak in Fig.\ 4 carries about
half quantized topological charge by performing area integral
around the sharp peak, while the rest of the area almost has no
contribution to the Chern number $C_{sc}$. With further increasing
$V_{R}$, these half quantized topological charges will mix and
merge with the ones coming from the upper energy band (with
opposite signs), when the band gap disappears and the two energy
bands touch. For a pure sample, a step jump of the $C_{sc}$ from
$2$ to $0$ at a critical Rashba coupling $V_{R}^{c}\simeq 0.33t$
is observed, where the energy band exactly collapses,
corresponding to a quantum phase transition into a metallic state.
The phase boundary for the topological quantized $C_{sc}=2$ QSHE
state has also been examined in the presence of disorder, where we
find that the number of edge channels $N_{ed}$ (see Fig.\ 3b) has
one to one correspondence to the spin Chern number $C_{sc}$, and
thus we identify the CNM as the topological origin of the
nontrivial QSHE.

To summarize, we have numerically studied several distinctive
properties of 2D honeycomb lattice model for the QSHE. We have
identified a pair of ideal conducting edge states within the bulk
band gap in an open system, which carry spin polarization and can
pump spins from one edge to the other upon an adiabatic insertion
of magnetic flux into a cylindric sample. These edge states can be
detected by using a Corbino disk of graphene with a circular
gate~\cite{explaugh}. Then we have established the relation of the
QSHE and the topology of the bulk states characterized by a spin
Chern number $C_{sc}=\pm 2$. The spin Chern number $C_{sc}$
remains precisely quantized in the QSHE phase until the band gap
collapses at strong disorder or Rashba coupling, where the edge
states disappear simultaneously.

\textbf{Acknowledgment:} D.N.S. would like to thank D. Arovas, B.
A. Bernevig, and S. -C. Zhang for stimulating discussions. This
work is supported by ACS-PRF 41752-AC10,  the
NSF grant/DMR-0307170 (DNS) and NSF (under MRSEC
grant/DMR-0213706) at the Princeton Center for Complex Materials
(FDMH), grants from NSFC (ZYW), and a grant from the Robert A.
Welch Foundation under Grant No. E-1146 (LS).


\begin{thebibliography}{99}
\bibitem{wen} X.G. Wen and Q.Niu, Phys. Rev. B \textbf{41}, 9377
(1990).

\bibitem{qheb} For a review see, \textit{The quantum Hall effect},
edited by R. E. Prange and S. M. Girvin (Springer-Verlag, Berlin,
1990).

\bibitem{thouless} D.~J. Thouless \emph{et al.}, Phys. Rev. Lett.
\textbf{49}, 405 (1982).


\bibitem{berry} M. V. Berry, Proc. R. Soc. Lond. A. $\mathbf{392}$, 45
(1984).

\bibitem{niu} Q. Niu \emph{et al.}, Phys. Rev. B \textbf{31},
3372 (1985).

\bibitem{haldane} F. D. M. Haldane, Phys. Rev. Lett. \textbf{93},
206602 (2004).

\bibitem{donna1} D. N. Sheng \emph{et al.}, Phys. Rev. Lett.
\textbf{90}, 256802 (2003).

\bibitem{haldane88} F. D. M. Haldane, Phys. Rev. Lett. \textbf{61},
2015 (1988).

\bibitem{kane} C.L. Kane and E.J. Mele, Phy. Rev. Lett. \textbf{95},
146802 (2005); \textit{ibid.} \textbf{95}, 226801 (2005).

\bibitem{lisheng} L. Sheng, D. N. Sheng, C. S. Ting, and F. D. M.
Haldane, Phy. Rev. Lett. \textbf{95}, 136602 (2005).


\bibitem{zhang} X.-L. Qi \emph{et al.}, cond-mat/0505308; B. A.
Bernevig and S. -C. Zhang, cond-mat/0504147; C. Wu \emph{et
al.}, cond-mat/0508273.

\bibitem{intrinsic} S. Murakami, N. Nagaosa, and S. C. Zhang, 
Science {\bf 301}, 1348 (2003);  J. Sinova \emph{et al.}, Phys. Rev.
Lett.  \textbf{92}, 126603 (2004).

\bibitem{moore} C. Xu and J. E. Moore, 
Phys. Rev. B {\bf 73}, 045322 (2006).

\bibitem{laugh} R.~B. Laughlin, Phys. Rev. Lett. \textbf{50}, 1395
(1983).


\bibitem{graph} K. S. Novoselov $et$ $al.$, Science {\bf 306}, 666
(2004).

\bibitem{donnarashba} D. N. Sheng \emph{et al.},
Phys. Rev. B \textbf{72}, 153307 (2005).



\bibitem{donna2} D. N. Sheng, L. Balents, and  Z. Wang, Phys. Rev. Lett.
\textbf{91}, 116802 (2003).

\bibitem{arou} D.~P. Arovas \emph{et al.,} Phys. Rev. Lett.
\textbf{60}, 619 (1988).

\bibitem{explaugh} V. T. Dolgopolov \emph{et al.}, Phys. Rev. B
\textbf{48}, 8480 (1993).

\end{thebibliography}
\end{document}